\documentclass[aps,twocolumn,prb,preprintnumbers,amsmath,amssymb]{revtex4}
\usepackage{graphicx, bm}
\usepackage{dcolumn}
\usepackage{float}
\usepackage{latexsym}
\usepackage{amsmath}
\usepackage{graphics}
\usepackage{amssymb}
\usepackage{layout}
\usepackage{verbatim}
\usepackage{amsfonts,epsfig}
\usepackage{color}
\usepackage{multirow}
\usepackage{setspace}

\newcommand{\beqnar}{\begin{eqnarray}}
\newcommand{\eeqnar}{\end{eqnarray}}

\newcommand{\beq}{\begin{equation}}
\newcommand{\eeq}{\end{equation}}

\begin{document}
\title{Two-dimensional metal-insulator-transition as a potential fluctuation driven semiclassical transport phenomenon}
\author{S. Das Sarma$^1$, E. H. Hwang$^{1,2}$, and Qiuzi Li$^1$}
\affiliation{$^1$Condensed Matter Theory Center, Department of Physics, University of Maryland, College Park, Maryland 20742\\
             $^2$SKKU Advanced Institute of Nanotechnology and Department of Physics,
Sungkyunkwan University, Suwon 440-746, Korea}
\date{\today}
\begin{abstract}
We theoretically consider the carrier density tuned (apparent) two-dimensional (2D) metal-insulator-transition (MIT) in semiconductor heterostructure-based 2D carrier systems as arising from a classical percolation phenomenon in the inhomogeneous density landscape created by the long-range potential fluctuations induced by random quenched charged impurities in the environment. The long-range Coulomb disorder inherent in semiconductors produces strong potential fluctuations in the 2D system where a fraction of the carriers gets trapped or classically localized, leading to a mixed 2-component semiclassical transport behavior at intermediate densities where a fraction of the carriers is mobile and another fraction immobile. At high carrier density, all the carriers are essentially mobile whereas at low carrier density all the carriers are essentially trapped since there is no possible percolating transport path through the lake-and-mountain inhomogeneous potential landscape. The low-density situation with no percolation would mimic an insulator whereas the high-density situation with allowed percolating paths through the lake-and-mountain energy landscape would mimic a metal with the system manifesting an apparent MIT in between. We calculate the transport properties as a function of carrier density, impurity density, impurity location, and temperature using a 2-component (trapped and mobile carriers) effective medium theory.  Our theoretically calculated transport properties are in good qualitative agreement with the experimentally observed 2D MIT phenomenology in 2D electron and hole systems. We find a high (low) density metallic (insulating) temperature-dependence of the 2D resistivity, and an intermediate-density crossover behavior which could be identified with the experimentally observed 2D MIT. The calculated density- and temperature-dependent resistivity in our theory mimics the phenomenology of 2D MIT experiments with reasonable parameter values for the background disorder.
\end{abstract}


\maketitle

\section{Introduction}

\begin{figure}[htb]
\begin{center}
\includegraphics[width=0.99\columnwidth]{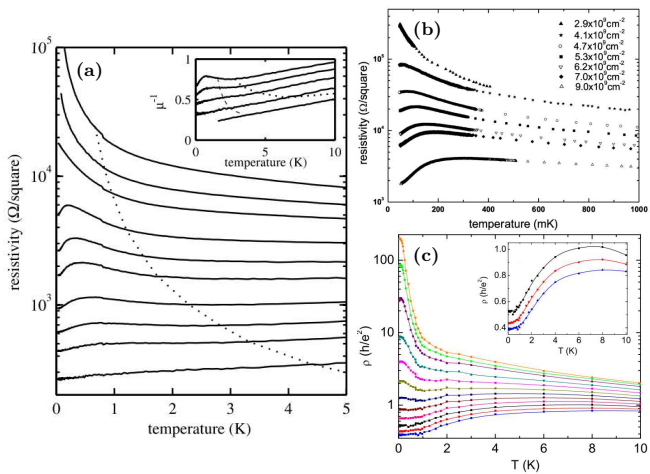}
  \caption{(Color online). (a) Temperature dependence of the observed resistivity for 2D electrons in GaAs heterostructure. Adapted from Ref.~[\onlinecite{lilly2003}]. (b)  Temperature dependence of the observed resistivity for p-type GaAs quantum well system. Adapted from Ref.~[\onlinecite{manfra2007}].  (c) Experimental resistivity as a function of temperature for Si MOSFETs. Adapted from Ref.~[\onlinecite{tracy2009}]. }
\label{fig1}
\end{center}
\end{figure}

The carrier density-tuned 2D MIT phenomenon is ubiquitous in semiconductor heterostructures at low temperatures\cite{Ando_RMP82,abrahams2001,kravchenko2004,dassarma2005,Spivak_RMP10,dassarma2010}. At ``higher" carrier density the measured resistivity manifests ``metallic" temperature dependence whereas at ``lower" carrier density the 2D resistivity manifests ``insulating" temperature dependence with a complex density-dependent crossover behavior in the ``intermediate" density regime where the system makes a ``transition" from being a high-density effective metal to a low-density effective insulator. In Fig.~\ref{fig1} we show, purely to provide a context for the current theoretical work, a set of representative 2D MIT experimental transport data\cite{lilly2003,manfra2007,tracy2009} for 2D electrons in Si MOSFETs, 2D electrons in GaAs heterostructures, and 2D holes in GaAs quantum wells, taken from the experimental publications of several different groups over the years. (These three systems are by far the most extensively studied experimental systems in the 2D MIT studies over the last 15 years\cite{jiang_apl1533,kravchenko1994,popovi_prl97,simmons_PRL98,Hanein_PRB98,Henein_PRL98,yoon_PRL99,Mills_PRL99,ilani2000,ilani2001,
lewalle_PRB02,lilly2003,leturcq_PRL03,leturcq_SPIE04,dassarma2005b,allison2006,manfra2007,anissimova_Nature07,laikeji_PRB07,adam2008,
tracy2009,qiurichard_PRB11}.)

The current theoretical work presented in this paper describes a simple and intuitively appealing semiclassical model as a plausible explanation for the 2D MIT phenomenon. After developing the model in depth, we carry out numerical calculations for the density and temperature dependent 2D resistivity which can be compared with the experimental observations (Fig.~\ref{fig1}). The qualitative agreement between our theoretical results and the experimental 2D MIT data is prima facie evidence  that our proposed physical mechanism is likely to be playing a role (perhaps even a major role) in the underlying physics of 2D MIT phenomena\cite{jiang_apl1533,kravchenko1994,popovi_prl97,Henein_PRL98,
ilani2000,ilani2001,lilly2003,leturcq_PRL03,leturcq_SPIE04,dassarma2005b,tracy2006,manfra2007,
laikeji_PRB07,adam2008,tracy2009,qiurichard_PRB11}. We cannot, however, rule out (certainly, not conclusively) the possibility of alternate physical mechanisms also playing some role in the 2D MIT phenomena since our model is based only on the single physical mechanism described in this work.

The physical mechanism we consider is disorder-induced density inhomogeneity necessarily present in semiconductors at low carrier densities where screening effects are sufficiently weak  so that the long-range Coulomb disorder introduced by random quenched charged impurities invariably present in  the semiconductor environment leads to strong potential fluctuations in the system\cite{gold_prb91,Coleridge_PRB97,wilamowski_PRL01}. These potential fluctuations are screened out at high carrier densities, but at low carrier densities they create highly inhomogeneous ``lakes-and-hills" type potential landscape in which the carriers move around within the physical sample\cite{fstern_prb74}. In addition to creating the lakes-and-hills (or equivalently ``valleys-and-mountains") inhomogeneous potential landscape, the charged impurities, of course, also cause momentum scattering of the carriers, leading to the measured resistivity. Since 2D MIT typically is a low-density phenomenon where screening is weak, the ``lakes-and-hills" potential landscape is expected to play a key role in the phenomenon.

The key ingredient of our model is that  we treat the carrier system as an effective 2-component (or equivalently, 2-phase) system: bound and unbound carriers or equivalently, trapped (i.e. immobile) and mobile (i.e. free) carriers (we use the latter terminology throughout), where all classically trapped carriers in the potential well (or valley) regions are assumed to be localized without contributing to the conductivity ($\equiv$ the inverse of resistivity) at zero temperature ($T=0$). Thus, the ``trapped" carriers in the classically forbidden region (with $E \leq 0$) do not contribute to the $T=0$ conductivity (or equivalently have infinite resistivity at $T=0$), and the $T=0$ conductivity is determined entirely by the mobile carriers (with $E \geq 0$) which are classically free. Our basic 2-component transport model is thus a classical (or semiclassical) model with the carriers being divided into bound or trapped or immobile carriers and unbound or free or mobile carriers because of the ``lakes-and-hills" potential fluctuation landscape induced by the random charged impurities. It is obvious that in the absence of Anderson localization (which we neglect) and in the absence of background potential fluctuations (or for very small fluctuations) all the carriers contribute to the $T=0$ ``metallic" conductivity since there are no effective potential wells or barriers to trap the carriers -- the same is essentially true when the net carrier density is very high so that the chemical potential or the Fermi energy is very large compared with the typical magnitude of the potential fluctuations. On the other hand, when the Fermi energy is very small (i.e. low carrier density), most of the carriers are immobile leading to exponentially small (and ``insulating" or activated) conductivity.

We assume that the mobile carriers undergo standard disorder scattering limited diffusive transport and as such can be treated by the semiclassical Boltzmann transport theory (taking into account the screening of the Coulomb disorder by the carriers themselves so that the resistive scattering mechanism is the finite-temperature screened Coulomb disorder arising from the quenched random charged impurities\cite{stern1980,dassarma1999,dassarma2000,dassarma2004}). At finite temperatures (but not at $T=0$), the trapped carriers also contribute to the conductivity through the activation process as the bound carriers can be thermally excited over the disorder-induced potential fluctuations to become effectively mobile. (We neglect all quantum tunneling and phonon-assisted hopping effects.) Thus, there are two independent transport channels in the problem: Diffusive transport by the mobile electrons/holes and activated transport by the trapped electrons/holes. We use a 2-component effective medium theory (EMT)\cite{kirkpatrick1973,qzli_PRB11,Hwang_InsuPRB2010,qzlidbo_PRB12,qiuzi_PRB12TI}, the two components being the fractions of trapped (``activated transport" ) and mobile (``diffusive transport") carriers, to describe the transport behavior in the system. Obviously, high-density transport, when the fraction of mobile carriers is very high, would appear diffusive and metallic, and low-density transport, when the fraction of trapped carriers is very high, would appear activated insulating with a crossover at some disorder-dependent intermediate carrier density.

\begin{figure}[htb]
\begin{center}
\includegraphics[width=0.99\columnwidth]{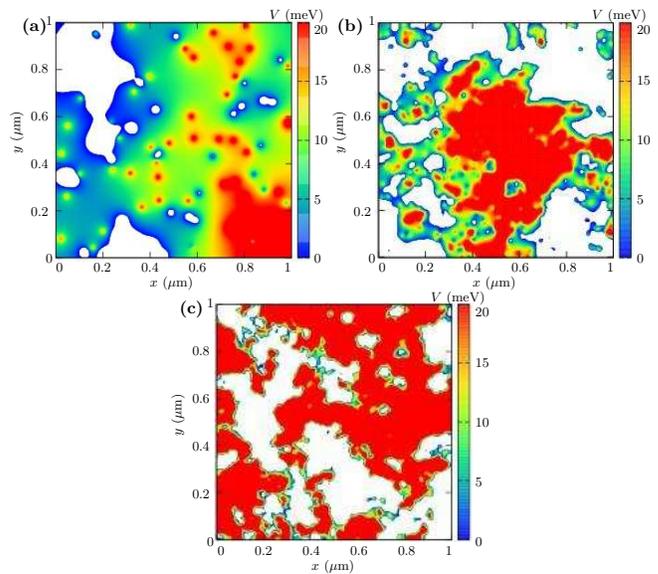}
  \caption{(Color online). 2D density plot of our simulated $V (\mathbf{r})$ measured from $E_F$ for randomly distributed impurities with $z_0 = 10$ nm and $\kappa = 10$. (a) $n_i = 10^{10}$ cm$^{-2}$. (b) $n_i = 10^{11}$ cm$^{-2}$.  (c) $n_i = 10^{12}$ cm$^{-2}$. Since this figure is just for schematic purposes, we do not include screening effects in our simulations. }
\label{fig2}
\end{center}
\end{figure}

\begin{figure}[htb]
\begin{center}
\includegraphics[width=0.99\columnwidth]{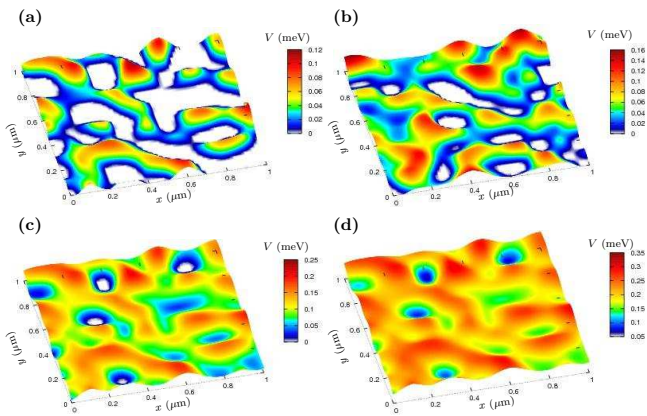}
  \caption{(Color online). Schematic diagram of $V (\mathbf{r})$  for $s=0.05$ meV to show the lakes-and-hills inhomogeneous potential landscape. (a) $E_F = 0.01$ meV. (b) $E_F = 0.05$ meV. (c) $E_F = 0.1$ meV. (d) $E_F = 0.2$ meV.  }
\label{fig3}
\end{center}
\end{figure}

We emphasize that our basic 2-component (trapped and mobile carriers) model is classical as is our 2-component EMT transport calculation\cite{kirkpatrick1973} in the sense that quantum interference (i.e. localization) effects are neglected. We calculate the carrier resistivity itself using the semiclassical Boltzmann transport theory which treats carrier momentum scattering by disorder in a quantum mechanical manner. In this sense, our theory is semiclassical. All quantum tunneling and quantum interference (and electron-electron interaction beyond screening) effects are ignored in our theory since the problem becomes intractable otherwise. Also, the 2-component classical transport model is inconsistent with quantum tunneling and interference (since in the presence of tunneling, the 2-component concept does not apply), and our goal here is to study the physically-motivated semiclassical 2-component transport model as completely as possible without the very difficult additional complications of quantum tunneling and interference. The basic idea here is that the strong potential fluctuations in the semiconductor system arising from random charged impurity disorder and the associated lakes-and-hills energy landscape necessarily lead to the zeroth order carrier transport properties being dominated by the 2-component behavior through the diffusion of mobile carriers and the activation of trapped carriers with all other effects (tunneling, interference, interaction) being weak perturbations in the strongly fluctuating highly inhomogeneous 2D system. The 2-component transport is relevant only when the Fermi energy of the carrier system and the typical scale of the potential fluctuations are comparable so that a highly inhomogeneous density and potential landscape dominates the transport properties (Figs.~\ref{fig2} and \ref{fig3}). This is the typical situation for 2D semiconductor systems in the presence of random charged impurity disorder where 2D MIT phenomena occur.

We mention that there is a very long history\cite{Herring_jap60,Kane_PR63,morgan_PR65,zallen1971,borisefros_spj71,kirkpatrick1973,Arnold_APL74,arnold1976} in the literature discussing transport in disordered semiconductor systems using the inhomogeneous 2-component model of trapped and mobile carriers in the lakes-and-hills landscape of potential fluctuations. The problem has a strong formal similarity to the problem of classical percolation in a highly inhomogeneous medium\cite{kirkpatrick1973,eggarter1970,efros_pss76,weinrib1982,bundle_JPA85,efros1989,stauffer_92,isichenko1992,he1998}, and occasionally the metal-insulator transition in electronic materials, particularly in disordered semiconductors, has been studied using the classical percolation approach\cite{batyev_JETP06}. The key physical point here is that the strong potential fluctuations induce strong density inhomogeneity with the electrons (or holes) forming spatial puddles separated by potential barriers\cite{pikus_spj89}. We mention here that a large number of 2D experimental\cite{lilly2003,manfra2007,tracy2009,jiang_apl1533,
ilani2000,ilani2001,lewalle_PRB02,leturcq_PRL03,leturcq_SPIE04,dassarma2005b,allison2006,
laikeji_PRB07,adam2008,qiurichard_PRB11,tracy2006} and theoretical\cite{qzli_PRB11,Hwang_InsuPRB2010,qzlidbo_PRB12,qiuzi_PRB12TI,efros1989,stauffer_92,isichenko1992,he1998,
batyev_JETP06,pikus_spj89,efros1988,fogler_PRB04,nixon1990,efros1993,shi2002} works in the literature have already suggested the 2-component percolation transport as the underlying mechanism for the 2D MIT phenomena.

In Section~\ref{sec:2} we describe our theory in details, providing the numerical results for the calculated 2D transport properties for various situations in Section~\ref{sec:3}. We provide a thorough discussion, emphasizing limitations of our theory and corrections to existing theoretical and experimental results in Section~\ref{sec:4}, concluding in Section~\ref{sec:5} with a discussion on the comparison with experiment and a summary.

\begin{figure}[htb]
\begin{center}
\includegraphics[width=0.99\columnwidth]{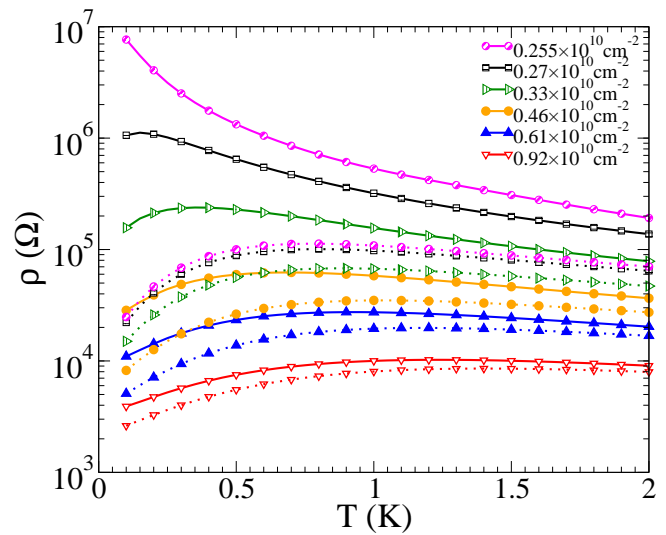}
  \caption{(Color online). Calculated $\rho (T)$ for various values of total carrier density $n$ for 2D holes in GaAs/AlGaAs quantum wells of width $a=200$ \AA \ assuming 2D interface impurity density of $1.5 \times 10^8$ cm$^{-2}$. The dotted and solid lines are for $s = 0$ and $s=0.05$ meV (corresponding to $n_t (T=0) = 0.25 \times 10^{10}$ cm$^{-2}$), respectively. The theoretical parameters in Figs.~\ref{fig4}-\ref{fig20} (i.e. all our presented results) are chosen to correspond to the experimental data in Ref.~[\onlinecite{manfra2007}].}
\label{fig4}
\end{center}
\end{figure}

\section{Theory}
\label{sec:2}
We use a minimal model to describe the 2D semiconductor system as a 2D electron (or hole) gas (2DEG) of 2D density $n$ characterized by an effective mass of $m$, a valley degeneracy factor of $g_v$, and a background effective dielectric constant $\kappa$ (which is taken as the average of the lattice dielectric constant for the semiconductor and the insulator defining the 2D heterostructure, i.e. Si-SiO$_2$, GaAs-AlAs, etc.). We assume a spin degenerate system with a spin degeneracy $g_s = 2$ throughout (except when discussing the effect of an in-phase magnetic field which could lift the spin degeneracy changing $g_s$ from two at zero magnetic field to one at high magnetic field). We neglect the thickness (in the $z$-direction perpendicular to the 2D layer) of the 2DEG in describing our theory and equations since this has only quantitative, but no qualitative, effect for the 2D MIT physics of interest in this work. Adding a finite 2D layer thickness effect is straightforward, and would only quantitatively modify the relative strength of the disorder in the theory. All our numerical results shown in the figures include the full effects of quasi-2D finite thickness in a realistic manner.

Disorder is included in our model through random charged impurities of concentration (i.e. 2D density) $n_i$ which are located a distance $z_0$ from the 2DEG in the $z$-direction perpendicular to the layer. We assume that $n_i$ and $z_0$ together define completely the resistive scattering of the carriers with the disorder potential being modeled by the screened Coulomb interaction.

\begin{figure}[htb]
\begin{center}
\includegraphics[width=0.99\columnwidth]{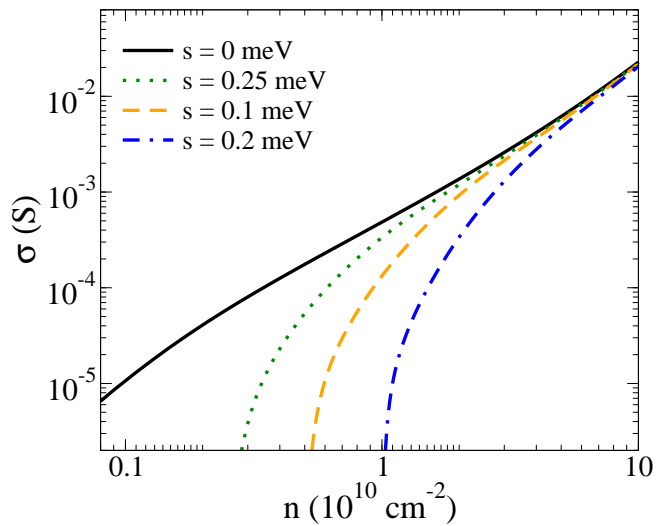}
  \caption{(Color online). Calculated $\sigma (n)$ at $T=50$ mK for different values of potential fluctuations $s$ for 2D holes in GaAs/AlGaAs quantum wells of width $a=200$ \AA \ assuming 2D interface impurity density of $1.5 \times 10^8$ cm$^{-2}$. }
\label{fig5}
\end{center}
\end{figure}

\begin{figure}[htb]
\begin{center}
\includegraphics[width=0.99\columnwidth]{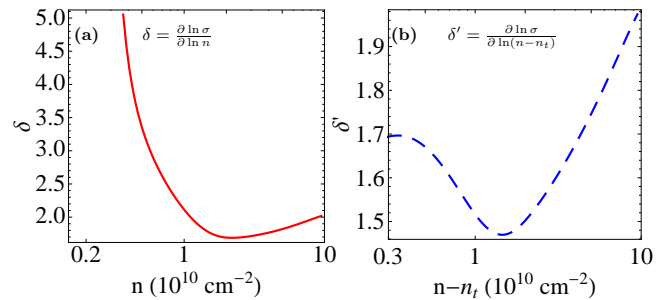}
  \caption{(Color online). Numerically calculated effective conductivity exponent ``$\delta$" at $T=50$ mK for $s = 0.05$ meV for 2D holes in GaAs/AlGaAs quantum wells of width $a=200$ \AA \ assuming 2D interface impurity density of $1.5 \times 10^8$ cm$^{-2}$. }
\label{fig6}
\end{center}
\end{figure}

Finally, we need to specify the potential fluctuations dominating the low-density behavior of the carrier system leading to the basic 2-component/2-phase model underlying our theoretical approach. In principle, the charged impurity distribution characterized by $n_i$ and $z_0$ and the 2D carrier system (characterized by $n$, $m$, $g_v$, $g_s$, $\kappa$) should suffice to define the disorder potential fluctuation distribution $P(V)$ through the self-consistent (and simultaneous) solutions of the Poisson equation for the charged impurity induced electrostatic potential
 $V (\mathbf{r})$ and the many-electron Schr\"{o}dinger equation for the response of the interacting 2D electron gas to the external disorder potential. Conceptually, one needs to use a general density functional approach where the disorder-induced potential $V (\mathbf{r})$  serves as the inhomogeneous potential and the ground state inhomogeneous carrier density $n(\mathbf{r})$ must minimize the appropriate (free) energy functional $E \{n(\mathbf{r})\}$. Given that the charged impurities are modeled by averaged quantities (e.g. $n_i$, $z_0$) since their precise locations are unknown, one must average over many different configurations of the impurity distribution after carrying out the full density functional self-consistent calculations for each configuration. Obviously, such a numerical density functional solution of the self-consistent problem (which would not only be nonlinear\cite{efros1988,fogler_PRB04}, but also nonlocal since there is no apriori reason to assume a local-density-approximation, LDA, to apply) is completely out of question because it will be computationally far too prohibitive (particularly since our goal is to obtain theoretical results for the 2D resistivity as a function of carrier density and temperature for different values of $n_i$, $z_0$, $m$, $g_v$, $\kappa$, etc.). In fact, even the ground state problem of obtaining just the inhomogeneous $n (\mathbf{r})$ for the disordered 2D system at $T=0$ in the presence of random quenched charged impurities in the background has only been attempted only a few times in the literature\cite{nixon1990,efros1993,shi2002,rossi2008,Rossi,polini2008,rossidassarma_11} within restrictive approximation schemes such as Hartree, LDA, and Thomas-Fermi theories. In spite of the simplified approximation schemes used in these theoretical works, it would be impossible to utilize any of these completely numerical nonlinear self-consistent approach to obtain the potential fluctuation distribution function for our purpose because combining such self-consistent calculations of disorder potential distributions with the requisite 2-component/2-phase transport calculations is simply beyond the computational power of currently available machines.

 Instead, we use the theoretically appealing approach of using a parameterized distribution function for the potential fluctuation function $P(V)$, which has recently been successful in developing a theoretical description for monolayer and bilayer graphene transport\cite{qzli_PRB11,qzlidbo_PRB12,Hwang_InsuPRB2010,qzli_PRL11,qiuzi_SCC12}  at low carrier densities around the Dirac point in the presence of electron-hole puddles induced by charged impurity disorder. In this theoretical approach, the disorder induced spatial potential fluctuation is approximated by a Gaussian function $P(V)$ which is assumed to be the same throughout the sample. Thus $V(\mathbf{r})$ at different $\mathbf{r}$ is assumed to be uncorrelated and described by a Gaussian function. The great advantage of this theoretical approach is its simplicity and conceptual clarity -- the potential distribution $P(V)$ can be completely characterized by a single energy ``$s$" which is the variance or the root-mean-square fluctuation in the impurity induced disorder potential\cite{mieghem_RMP92}. We assume, with no loss of generality, equal numbers of positive and negative random quenched point charges ($\pm e$) to be contributing to the impurity disorder so that there is no average (or net) potential contributed by disorder. Assuming uncorrelated random Poisson distribution in the impurity locations (all impurities located at random 2D positions in a layer $z_0$ away from the 2DEG) it is straightforward (but somewhat messy) to show that ``$s$" is related to the impurity and the 2DEG parameters (see Eq.~\eqref{eq:2} below) if linear Thomas-Fermi screening holds in the problem (which certainly does not hold at low carrier density). We do not, however, assume any relationship between $n_i$, $z_0$ on the one hand and $s$ on the other hand, taking $n_i$, $z_0$, and $s$ together to define a minimal $3-$ parameter description for the random disorder, where $s$ (relative to the Fermi energy $E_F$) defines the strength of the potential fluctuations and $n_i$, $z_0$ together gives the strength of diffusive resistive scattering of the mobile carriers. For $s=0$, all the carriers are mobile, and for infinite $s$ all the carriers are trapped.

 The probability $P(V)dV$ of finding the local electronic potential energy within a range $dV$ around $V$ is a Gaussian form:
\beq
P(V)=\dfrac{1}{\sqrt{2\pi s^2}} \exp(-V^2/2s^2)
\eeq
where $s$ denotes the standard deviation of potential fluctuations. Generally, larger $s$ means more impurity disorder (and more inhomogeneity) in the system. The potential fluctuation is directly related to the charged impurity density\cite{Stern_PR67,arnold1976,fstern_ss76}, and within the linear Thomas-Fermi screening theory:
\begin{equation}
s= \dfrac{e^2 \sqrt{\pi n_i}}{\sqrt{2} \kappa z_0 q_{TF}}
\label{eq:2}
\end{equation}
where $n_i$ is the density of charged impurity density, $\kappa$ is the arithmetic average of the semiconductor and insulator dielectric constant, $q_{TF}=\dfrac{g_s g_v m e^2}{\kappa \hbar^2}$ ($g_s$ and $g_v$ are the spin and valley degeneracy, respectively) is the Thomas-Fermi 2D linear screening wavevector, $m$ is the effective mass of the carriers, $z_0$ is the average distance of the charged impurity to the interface. We choose $s$ as a free parameter in the calculations (along with $n_i$ and $z_0$) below as explained above -- the reason being that linear screening in general breaks down at low carrier density.

The density of states $D(E)$ in the presence of potential fluctuations $P(V)$ is given in terms of the ideal 2D density of states $D_0$ in the absence of disorder:
\begin{eqnarray}
D(E) = \int _{-\infty }^E\dfrac{g_s g_v m}{2\pi \hbar^2} P(V) dV =\dfrac{D_0}{2}  \text{erfc}(-\dfrac{E}{\sqrt{2} s})
\end{eqnarray}
where $D_0$ is the ideal disorder-free 2D density of states given by:
\beq
D_0 = \dfrac{g_s g_v m}{2\pi \hbar ^2}
\label{eq:d0}
\eeq

At $T=0$, the carrier density in the band tail, i.e. the trapped or immobile carrier density, is given by the following formula:
 \begin{eqnarray}
n_t (T=0)  = D_0 \int _{-\infty }^{0}  \big[\dfrac{1}{2} \text{erfc}(-\dfrac{E}{\sqrt{2} s})\big]dE=D_0 \dfrac{s}{\sqrt{2 \pi }}
\label{Eq0:de}
\end{eqnarray}
Note that Eq.~\eqref{Eq0:de} defines the classically trapped fraction of the carrier density being in the classically forbidden negative energy states.

The total carrier density is given by (with $\mu$ being the finite-$T$ chemical potential with $E_F = \mu (T=0)$):
\beq
n = \int _{-\infty }^\infty D (E) \dfrac{dE}{\exp(\beta (E-\mu))+1}
\eeq
where $\mu$ is determined by the conservation of the carrier density:
\begin{eqnarray}
n &=& \int _{-\infty }^\infty D (E) \dfrac{dE}{\exp(\beta (E-\mu))+1} \nonumber
\\
\nonumber  \\
&=&\int _{-\infty }^{E_F} D (E)dE \ \ \ \ \text{at \ $T =0$} \nonumber
\\
\nonumber  \\
&=& D_0 \big[\dfrac{E_F}{2} \text{erfc}(-\dfrac{E_F}{\sqrt{2} s})+\dfrac{s}{\sqrt{2 \pi}} \exp(-\dfrac{E_F^2}{2s^2})\big]
\label{eq:totaln}
\end{eqnarray}

The trapped electron density at finite temperatures, $n_t(T)$  is determined by the potential fluctuation strength, but also depends on the temperature and is given by:
\begin{eqnarray}
n_t  = D_0 \int _{-\infty }^{0}  \big[\dfrac{1}{2} \text{erfc}(-\dfrac{E}{\sqrt{2} s})\big] \dfrac{dE}{\exp(\beta (E-\mu))+1}
\label{Eq:de}
\end{eqnarray}

The effective mobile or free carrier density, $n_m = n - n_t$, depends on the temperature, the Fermi energy $E_F$ and the standard deviation of potential fluctuation $s$. At zero temperature $T=0$, the effective mobile carrier density is given by:
\beq
n_m (T=0)=D_0 \big[\dfrac{E_F}{2} \text{erfc}(-\dfrac{E_F}{\sqrt{2} s})+\dfrac{s}{\sqrt{2 \pi}} \exp(-\dfrac{E_F^2}{2s^2})-\dfrac{s}{\sqrt{2 \pi}} \big]
\eeq
Thus, $n = n_m + n_t$ is the total carrier density with $n_m$, $n_t$ being the temperature-dependent fractions of mobile and trapped carrier densities. These are the two phases ($n_m$ and $n_t$) coexisting in the disordered 2DEG, and we must now develop a 2-component transport theory to obtain the 2D conductivity. The activated conductivity of the trapped immobile carriers is given by
\begin{eqnarray}
\sigma_a(V)=  \sigma  \exp(\beta (E_F-V))
\end{eqnarray}
where $\sigma = \dfrac{n e^2 \langle \tau\rangle}{m}$ with $\tau$ as the transport scattering time. We note that the variable range hopping should also be included in the conductivity of the trapped carriers, but we neglect it for two reasons. First, the variable range hopping is a phonon-assisted quantum tunneling phenomenon for the localized carrier, and our neglecting this is consistent with our neglect of all quantum tunneling processes in the theory. Second, including a hopping conduction would simply add more unknown parameters to the theory without adding any physical clarity.

For the mobile carriers we simply have the diffusive conductivity from the Boltzmann theory:
\beq
\sigma_d=  \dfrac{(n - n_t) e^2 \langle \tau\rangle}{m}=  \dfrac{n_m e^2 \langle \tau\rangle}{m}
\eeq
where the finite-temperature averaged $\langle \tau\rangle$ is given by\cite{dassarma2004,dassarma1999,dassarma2003}:
\begin{equation}
\langle \tau\rangle = \dfrac{\int d\epsilon D(\epsilon) \epsilon \tau(\epsilon)(-\partial f/\partial\epsilon)}{\int d\epsilon D(\epsilon) f(\epsilon)}
\end{equation}
Here we use $D(\epsilon)=D_0$ for the mobile carriers. $f(\epsilon)$ is the Fermi-Dirac distribution function:
\beq
f(\epsilon)=\dfrac{1}{1+e^{\beta (\epsilon-\mu_0)}}
\eeq
with $\mu_0$ as the finite temperature chemical potential for the homogeneous system, which is determined by:
\beq
\mu_0 = E_F + k_B T \log(1-e^{-\frac{E_F}{k_B T}})
\eeq
The energy-dependent scattering time is given by:
\begin{eqnarray}
\dfrac{\hbar}{\tau(\epsilon_{{\bf k}})}= 2\pi n_{i} \int \frac{d^2 k'}{(2\pi)^2}\Big|\dfrac{V(q)}{\varepsilon(q)}\Big|^2\times \left[1-\cos\theta\right]\delta(\epsilon_{\mathbf{k'}}-\epsilon_{\mathbf{k}})
\label{eq:mscatt}
\end{eqnarray}
where $V(q) = \frac{2 \pi e^2 }{\kappa q} e ^{-q z_0}$ is the scattering potential strength and $\varepsilon(q) = 1 - \frac{2 \pi e^2}{\kappa q} \Pi(q,T)$ is the finite-temperature 2D screening function\cite{stern1967,Ando_RMP82,ando_jpjsj82,dassarmaft2des_PRB86}.

We define the fractional occupancy as given by:
\beq
p=\int _{-\infty}^{E_F} P(V) dV   ,
\eeq
In the 2-component effective medium theory, we have the conductivity $\sigma_1$ and $\sigma_2$ for the two phases ``1" and ``2". In our case, phase 1 consists of mobile carrier:
\beq
\sigma_1= \sigma_d
\eeq
The conductivity ($\sigma_2 = \sigma_a$) of phase 2 corresponds to the activated conduction of the trapped carriers:
\begin{eqnarray}
\sigma_2 & = &\frac{1}{q}\int_{E_F}^{\infty}\sigma_a(V) P(V) dV
\nonumber \\
&=& \frac{\sigma}{2q} e^{
    \frac{\beta^2s^2}{2} +\beta E_F } {\rm erfc} \left (
    \frac{E_F}{\sqrt{2} s} + \frac{\beta s}{\sqrt{2}} \right) \nonumber \\
    &=& \sigma_a
\label{eq:sig2}
\end{eqnarray}
where $q = 1-p$.

The total conductivity is given by (according to the effective medium theory\cite{kirkpatrick1973,meir_PRB01})
\beq
\sigma = \dfrac{2p-1}{2}\Big[(\sigma_1-\sigma_2)+\sqrt{(\sigma_1-\sigma_2)^2+\dfrac{4}{(2p-1)^2}\sigma_1 \sigma_2}\Big]
\label{eq:sigmat}
\eeq
with $\sigma_1 = \sigma_d$ and $\sigma_2 = \sigma_a$.

In the next section, we give numerical results for $\rho = 1/\sigma$ in terms of system parameters.

\begin{figure}[htb]
\begin{center}
\includegraphics[width=0.99\columnwidth]{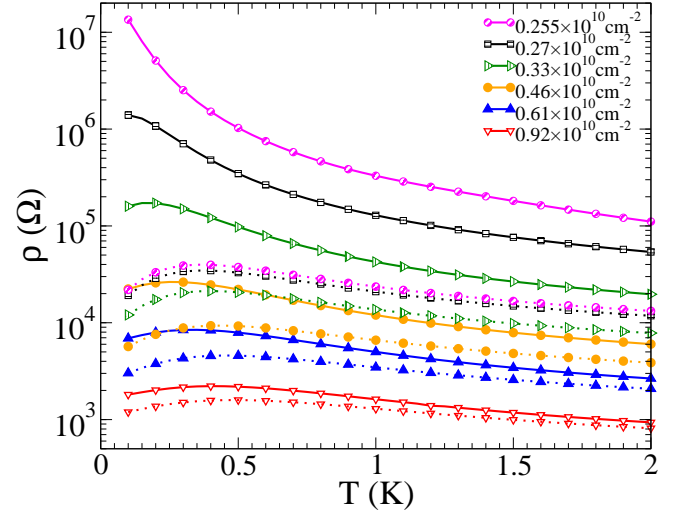}
  \caption{(Color online). Calculated $\rho (T)$ for various values of total carrier density $n$ for 2D holes in GaAs/AlGaAs quantum wells of width $a=200$ \AA \  assuming remote 2D impurity density of $3.5 \times 10^8$ cm$^{-2}$ and $z_0 = 300$ \AA. \ The dotted and solid lines are for $s = 0$ and $s=0.05$ meV (corresponding to $n_t (T=0) = 0.25 \times 10^{10}$ cm$^{-2}$), respectively. }
\label{fig7}
\end{center}
\end{figure}

\begin{figure}[htb]
\begin{center}
\includegraphics[width=0.99\columnwidth]{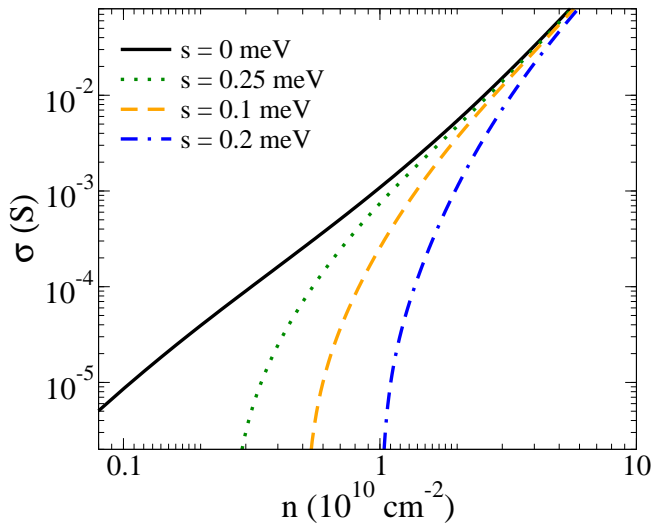}
  \caption{(Color online). Calculated $\sigma (n)$ at $T=50$ mK for different values of potential fluctuations $s$ for 2D holes in GaAs/AlGaAs quantum wells of width $a=200$ \AA \   assuming remote 2D impurity density of $3.5 \times 10^8$ cm$^{-2}$ and $z_0 = 300$ \AA.  }
\label{fig8}
\end{center}
\end{figure}

\begin{figure}[htb]
\begin{center}
\includegraphics[width=0.99\columnwidth]{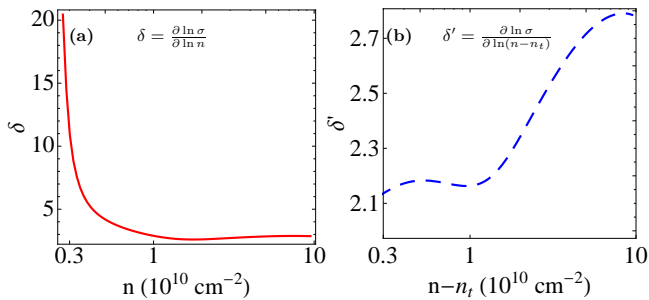}
  \caption{(Color online). Numerically calculated effective conductivity exponent ``$\delta$" at $T=50$ mK for $s = 0.05$ meV for 2D holes in GaAs/AlGaAs quantum wells of width $a=200$ \AA \   assuming remote 2D impurity density of $3.5 \times 10^8$ cm$^{-2}$ and $z_0 = 300$ \AA. }
\label{fig9}
\end{center}
\end{figure}

 Before presenting our numerical results for the carrier density and temperature dependent 2D resistivity in terms of $s$, $n_i$, $z_0$, we mention that in spite of the conceptual simplicity of the 2-component effective medium theory description of the 2-phase transport of the 2DEG in the presence of potential fluctuations, the results arise from a complex interplay of many distinct physical mechanisms which are not easy to grasp because of the large number of parameters defining the 2D MIT phenomena. First, there is a strong density and temperature dependence of $\sigma_d$, i.e. the diffusive conductivity of mobile carriers, itself arising from the variation in $q_{TF}/2k_F$ (``screening") and $k_B T/E_F$ (``degeneracy") as discussed already in details in the literature\cite{dassarma1999,dassarma2000,dassarma2003,dassarma2004}. This screening and degeneracy dependence of 2D ``metallic" transport is  included in our theory and is actually enhanced by the fact that $p<1$ by virtue of $s \neq 0$. Thus, the potential fluctuations enhance the temperature dependence of $\sigma_d$ (or $\sigma_1$) indirectly. More importantly, however, the activated conduction inherent in $\sigma_2$ ($=\sigma_a$) strongly enhances the temperature dependence of the conductivity, particularly at lower carrier densities (i.e. larger values of $s/E_F$) where the trapped carrier fraction is large. Since the two contributions to the conductivity (i.e. $\sigma_d$ and $\sigma_a$ or equivalently $\sigma_1$ and $\sigma_2$) have opposite temperature dependencies at low temperatures, it is possible for the system to exhibit a very weak temperature dependence at some intermediate carrier density mimicking a quantum critical density between a metal and an insulator whereas in reality, within our theory, it is simply a 2-phase crossover behavior where the 2D metallic phase at high density (and large $E_F$) consisting mainly of mobile/diffusive carriers is crossing over to a low density (and small $E_F$) 2D system consisting primarily of trapped/insulating 2D carriers. Our theory by construction is a crossover 2-component theory.

\section{Transport in the 2-component model}
\label{sec:3}
In providing the numerical results for the 2D transport properties in the 2-component effective medium theory approximation developed in section~\ref{sec:2} we mention that the quantitative results would depend strongly on the 2D material (since the relevant system parameters such as $q_{TF}$ and $E_F$ depend on materials parameters such $m$, $g_v$, and $\kappa$) as well as on the nature of the 2D impurity disorder. In particular, whether the dominant disorder arises from near impurities (relatively small values of $z_0$) or far impurities (large $z_0$) has a strong effect on the results. Another possibility we will consider is a background 3D impurity distribution in the 2D quantum well itself for which an integration would have to be carried out over a range of $z_0$ values and the theory of section~\ref{sec:2} has to be generalized for a finite 2D layer thickness (``$a$") which is straightforward to do assuming a square-well quantum 2D confinement in the $z-$direction with a layer thickness of $a$.

We provide our numerical results only for 2D GaAs-AlGaAs hole system (corresponding to Ref.~[\onlinecite{manfra2007}]) since typically this system exhibits the strongest 2D MIT behaviors although we have produced transport results for many different 2D systems used in various experimental studies obtaining qualitatively similar results.  Below we discuss and present our numerical transport results corresponding to Ref.~[\onlinecite{manfra2007}] by considering individual models for impurity disorder arising from 2D interface impurities (\ref{subsec:inter}), remote 2D impurities (\ref{subsec:remo}),  background 3D impurities (\ref{subsec:back}), and remote 3D impurities (\ref{subsec:remo3D})  We mention that in reality, all four disorder mechanisms are likely to be present with varying quantitative magnitudes.

\subsection{Interface 2D impurity model}
\label{subsec:inter}
We show in  Fig.~\ref{fig4} our calculated $\rho (T)$ for various values of total carrier density $n$ for $s=0.05$ meV for 2D holes in GaAs/AlGaAs quantum wells of width $a=200$ \AA \ assuming 2D interface impurity density of $1.5 \times 10^8$ cm$^{-2}$. This is a rather clean system corresponding to $n_t (T=0) = 0.25 \times 10^{10}$ cm$^{-2}$. The system corresponds to the 2D hole system studied in Ref.~[\onlinecite{manfra2007}]. In Fig.~\ref{fig5} we show the corresponding $\sigma(T)$ plots (with $\sigma \equiv \rho^{-1}$ by definition) at $T=50$ mK for different values of $s=0, \ 0.05, \ 0.1$, and $0.2$ meV. Results shown in Figs.~\ref{fig4} and \ref{fig5} indicate that the 2-component effective medium approximation, assuming conduction by diffusive mobile carriers and activated trapped carriers, qualitatively reproduce the observed features of the 2D MIT phenomena. In particular, $\rho(T)$ mimics a metal-insulator transition around $n_t$, and the critical density for the 2D MIT crossover depends strongly on the magnitude of the potential fluctuations with the crossover ``critical" density increasing with increasing ``$s$".

\begin{figure}[htb]
\begin{center}
\includegraphics[width=0.99\columnwidth]{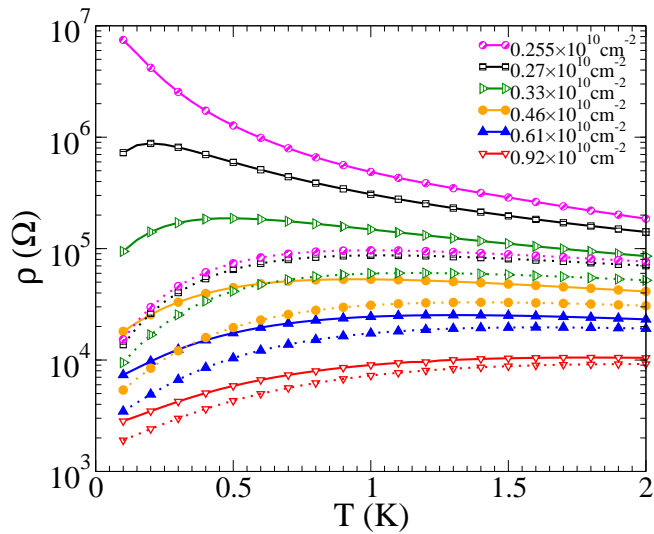}
  \caption{(Color online). Calculated $\rho (T)$ for various values of total carrier density $n$  for 2D holes in GaAs/AlGaAs quantum wells of width $a=200$ \AA. \  The impurities are distributed randomly throughout the quantum well with 3D average density $7.5 \times 10^{13}$ cm$^{-3}$. The dotted and solid lines are for $s = 0$ and $s=0.05$ meV (corresponding to $n_t (T=0) = 0.25 \times 10^{10}$ cm$^{-2}$), respectively.}
\label{fig10}
\end{center}
\end{figure}

\begin{figure}[htb]
\begin{center}
\includegraphics[width=0.99\columnwidth]{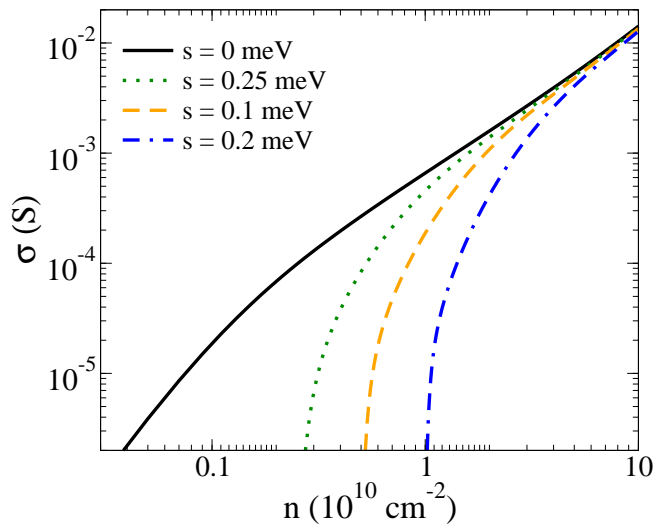}
  \caption{(Color online). Calculated $\sigma (n)$ at $T=50$ mK for different values of potential fluctuations $s$ for 2D holes in GaAs/AlGaAs quantum wells of width $a=200$ \AA. \   The impurities are distributed randomly throughout the quantum well with 3D average density $7.5 \times 10^{13}$ cm$^{-3}$.  }
\label{fig11}
\end{center}
\end{figure}

\begin{figure}
\begin{center}
\includegraphics[width=0.99\columnwidth]{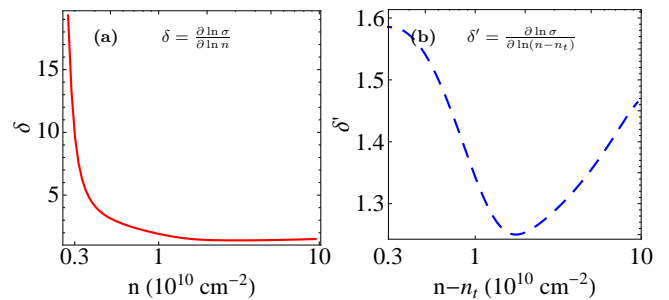}
  \caption{(Color online). Numerically calculated effective conductivity exponent ``$\delta$" at $T=50$ mK for $s = 0.05$ meV for 2D holes in GaAs/AlGaAs quantum wells of width $a=200$ \AA. \   The impurities are distributed randomly throughout the quantum well with 3D average density $7.5 \times 10^{13}$ cm$^{-3}$. }
\label{fig12}
\end{center}
\end{figure}

In Fig.~\ref{fig6} we plot the numerical value of the effective conductivity exponent ``$\delta$" defined as either $\delta = \frac{\partial \ln \sigma}{ \partial \ln n}$ or $\delta' = \frac{\partial \ln \sigma}{ \partial \ln (n- n_t)}$ as a function of carrier density for $s=0.05$ meV, showing that $\delta$ and $\delta'$ are also consistent with experimental findings\cite{manfra2007}. In particular, $\delta$ itself becomes very large as density decreases but $\delta' \sim 1.5$ consistent with a percolation picture\cite{leturcq_PRL03,dassarma2005b,manfra2007,tracy2009}.

We emphasize that there is no point in demanding strict quantitative agreement between theory and experiment since $n_i$, $z_0$, and $s$ are all unknown experimentally (and a minimal model involving only three parameters is unlikely to capture all the complexity of the actual disorder in the realistic samples).

\subsection{Remote 2D impurity model}
\label{subsec:remo}
In Figs.~\ref{fig7}-\ref{fig9} we show our calculated results for remote 2D impurity scattering, again for 2D holes in GaAs/AlGaAs quantum wells. Results are very similar for the near impurity case (Figs.~\ref{fig4}-\ref{fig6}) except that the temperature dependence in the metallic phase for $n  \gg n_t$ is weaker than the interface impurity case of the last section because $2 k_F$ scattering is suppressed in the remote impurity case.

\subsection{Background 3D impurity model}
\label{subsec:back}
Here the impurities are distributed randomly throughout the quantum well with some 3D average impurity density--often in very high mobility ultrapure  2D systems, unintentional 3D background charged impurities at some low concentration dominate transport properties. In Figs.~\ref{fig10}-\ref{fig12} we show our numerical results for transport limited by 3D background impurities. Again, the results are qualitatively similar to those in Figs.~\ref{fig4}-\ref{fig9} except that the metallic temperature dependence of the high-density resistivity is somewhat stronger in this case since the impurities reside in the 2D layer itself.

\begin{figure}[htb]
\begin{center}
\includegraphics[width=0.99\columnwidth]{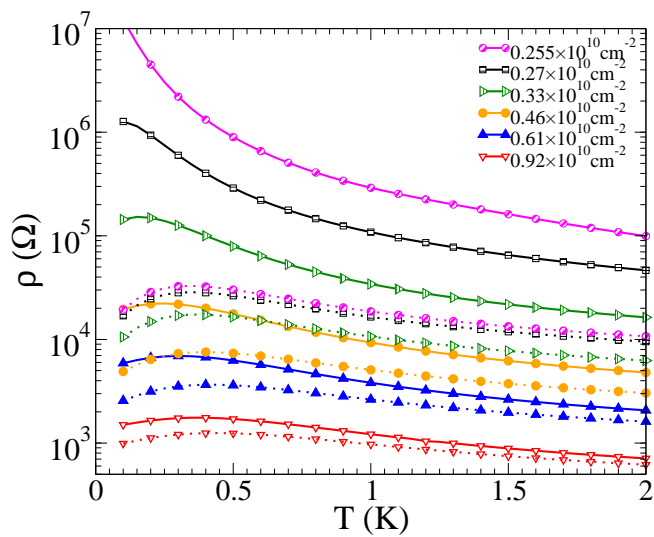}
  \caption{(Color online). Calculated $\rho (T)$ for various values of total carrier density $n$  for 2D holes in GaAs/AlGaAs quantum wells of width $a=200$ \AA. \  The 3D remote impurities are distributed randomly from $z_1 = 300$ \AA \ to $z_2 = 350$ \AA \  with 3D average density $7 \times 10^{14}$ cm$^{-3}$. The dotted and solid lines are for $s = 0$ and $s=0.05$ meV (corresponding to $n_t (T=0) = 0.25 \times 10^{10}$ cm$^{-2}$), respectively. }
\label{fig13}
\end{center}
\end{figure}

\begin{figure}[htb]
\begin{center}
\includegraphics[width=0.99\columnwidth]{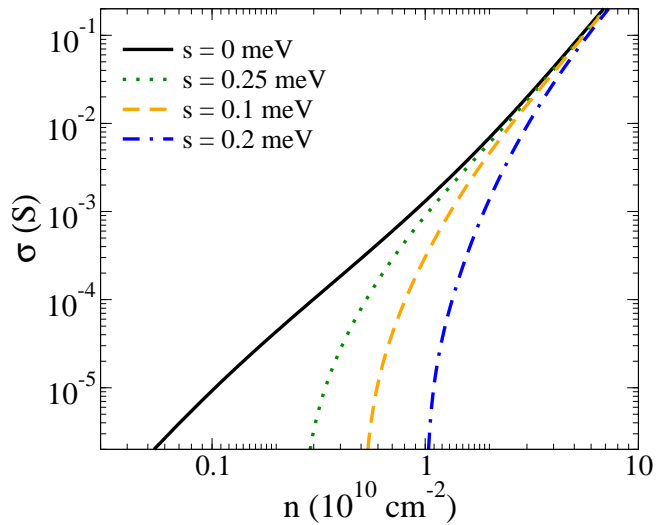}
  \caption{(Color online). Calculated $\sigma (n)$ at $T=50$ mK for different values of potential fluctuations $s$ for 2D holes in GaAs/AlGaAs quantum wells of width $a=200$ \AA. \  The 3D remote impurities are distributed randomly from $z_1 = 300$ \AA \ to $z_2 = 350$ \AA \  with 3D average density $7 \times 10^{14}$ cm$^{-3}$.}
\label{fig14}
\end{center}
\end{figure}

\begin{figure}
\begin{center}
\includegraphics[width=0.99\columnwidth]{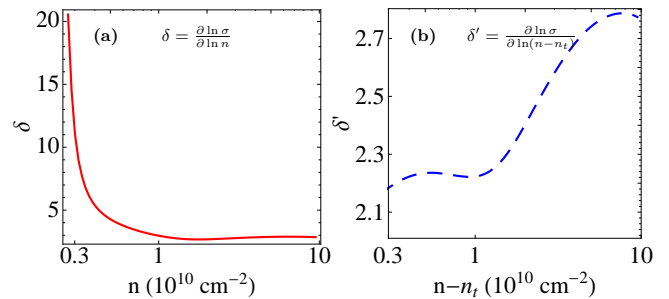}
  \caption{(Color online). Numerically calculated effective conductivity exponent ``$\delta$" at $T=50$ mK for $s = 0.05$ meV for 2D holes in GaAs/AlGaAs quantum wells of width $a=200$ \AA. \  The 3D remote impurities are distributed randomly from $z_1 = 300$ \AA \ to $z_2 = 350$ \AA \  with 3D average density $7 \times 10^{14}$ cm$^{-3}$.}
\label{fig15}
\end{center}
\end{figure}

\subsection{Remote 3D impurity model}
\label{subsec:remo3D}
Here the charged impurities are distributed randomly in three dimensions over a remote spacer layer located away from the 2D quantum well. These results are shown in Figs.~\ref{fig13}-\ref{fig15}, and are similar to the results in Figs.~\ref{fig4}-\ref{fig12} except for quantitative differences.

We conclude this section on our numerical results by emphasizing that we can easily produce results for transport limited by different combinations of impurity scattering (i.e. combinations of 2D near and far scatters and/or of 3D background and remote scatters), but the qualitative behavior would remain the same. Therefore, we do not see any compelling reason to present additional numerical results including all four disorder models together which will in principle involve 12 unknown disorder parameters. For the same reason we refrain from providing numerical transport results for other 2D systems since the qualitative physics would remain the same and the results would look very similar with obvious quantitative differences.

\begin{figure}[htb]
\begin{center}
\includegraphics[width=0.99\columnwidth]{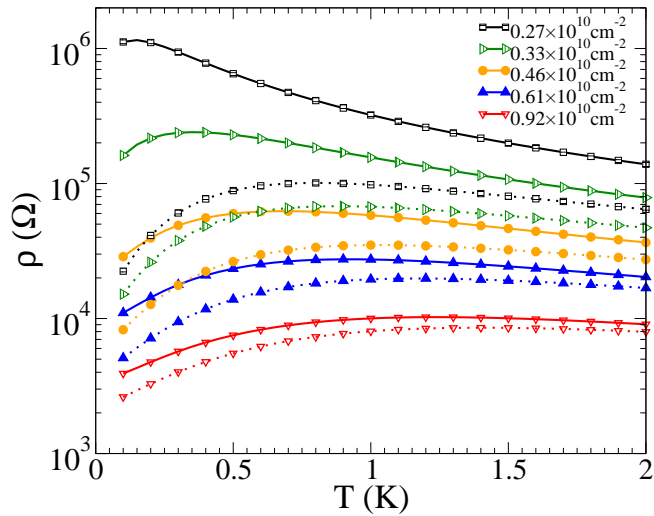}
  \caption{(Color online).  Calculated $\rho (T)$ for various values of total carrier density $n$  for 2D holes in GaAs/AlGaAs quantum wells of width $a=200$ \AA. \  The 2D interface impurity density is $1.5 \times 10^8$ cm$^{-2}$. The 3D remote impurities are distributed randomly from $z_1 = 1600$ \AA \ to $z_2 = 1650$ \AA \  with 3D average density $7 \times 10^{14}$ cm$^{-3}$. The dotted and solid lines are for $s = 0$ and $s=0.05$ meV (corresponding to $n_t (T=0) = 0.25 \times 10^{10}$ cm$^{-2}$), respectively. }
\label{fig16}
\end{center}
\end{figure}

\begin{figure}[htb]
\begin{center}
\includegraphics[width=0.99\columnwidth]{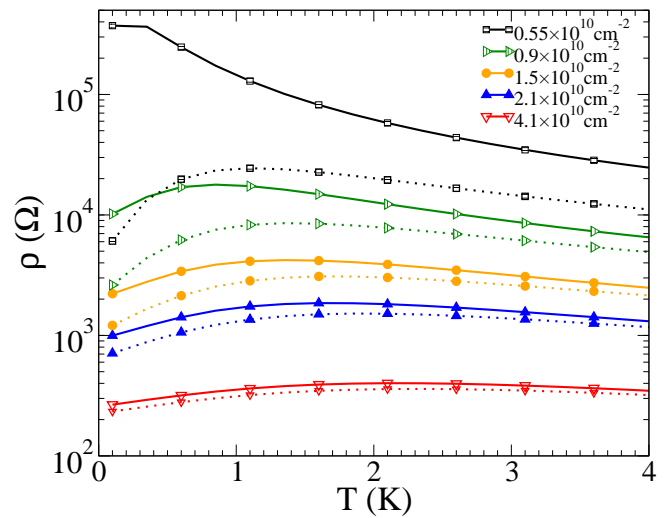}
  \caption{(Color online). Calculated $\rho (T)$ for various values of total carrier density $n$ for 2D holes in GaAs/AlGaAs quantum wells of width $a=200$ \AA. \  The 2D interface impurity density is $1.5 \times 10^8$ cm$^{-2}$. The 3D remote impurities are distributed randomly from $z_1 = 1600$ \AA \ to $z_2 = 1650$ \AA \  with 3D average density $7 \times 10^{14}$ cm$^{-3}$. The dotted and solid lines are for $s = 0$ and $s=0.1$ meV (corresponding to $n_t (T=0) = 0.5 \times 10^{10}$ cm$^{-2}$), respectively.  }
\label{fig17}
\end{center}
\end{figure}

\begin{figure}[htb]
\begin{center}
\includegraphics[width=0.99\columnwidth]{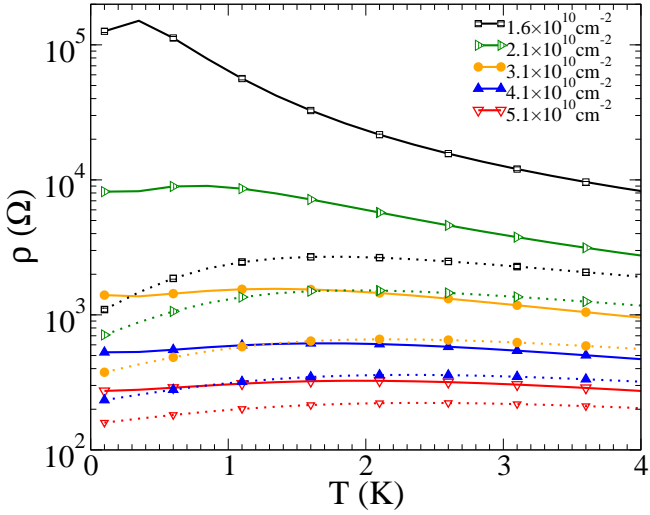}
  \caption{(Color online). Calculated $\rho (T)$ for various values of total carrier density $n$  for 2D holes in GaAs/AlGaAs quantum wells of width $a=200$ \AA. \  The 2D interface impurity density is $1.5 \times 10^8$ cm$^{-2}$. The 3D remote impurities are distributed randomly from $z_1 = 1600$ \AA \ to $z_2 = 1650$ \AA \  with 3D average density $7 \times 10^{14}$ cm$^{-3}$. The dotted and solid lines are for $s = 0$ and $s=0.3$ meV (corresponding to $n_t (T=0) = 1.5 \times 10^{10}$ cm$^{-2}$), respectively.  }
\label{fig18}
\end{center}
\end{figure}

Instead, in Figs.~\ref{fig16}-\ref{fig18} we show the effect of having different values of potential fluctuations (by varying ``$s$") on our calculated 2D MIT phenomenology. These figures  show that increasing ``$s$", i.e. having stronger disorder, suppresses the metallic temperature dependence, explaining why the strong metallicity (i.e. strong metallic temperature dependence of the 2D resistivity) necessitates having cleaner samples which presumably have smaller values of ``$s$".

\section{Discussion}
\label{sec:4}
The main point established through our  numerical results (Figs.~\ref{fig4}-\ref{fig18}) presented in this paper is that a theory based on a 2-component transport mechanism involving a 2-phase coexistence of both mobile (diffusive) and trapped (activated) carriers in the inhomogeneous ``lakes-and-hills" landscape of disorder-induced potential fluctuations provides a reasonable qualitative description of the phenomenology associated with the experimental 2D MIT observations. In particular, the metal-to-insulator low density crossover and the associated temperature dependence of the resistivity in the experimental observations are similar to our theoretical results.

Below we provide some critical remarks in the context of a series of questions which might arise with respect to any theory purporting to provide a complete qualitative answer to the substantial (and long-standing) puzzle which goes by the name of 2D MIT. The answers to these questions are all qualitative, but our theory, as shown in Sec.~\ref{sec:3}, is capable of providing detailed quantitative transport results if the impurity disorder is known in quantitative depth (which is, of course, an impossibility in real laboratory samples).

\begin{figure}[htb]
\begin{center}
\includegraphics[width=0.99\columnwidth]{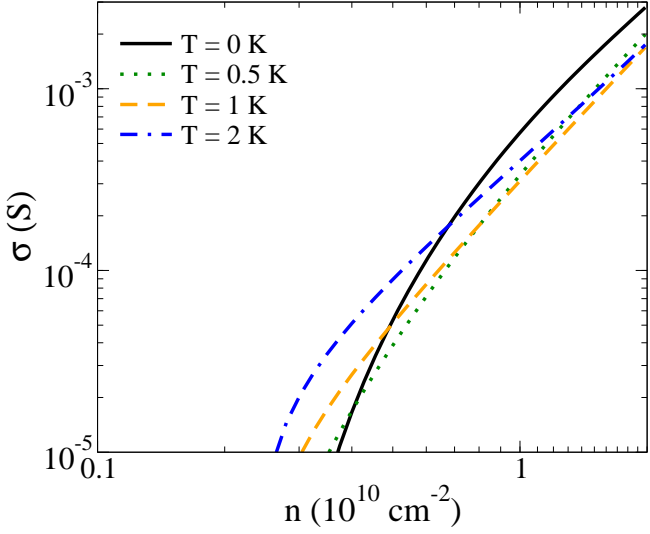}
  \caption{(Color online). Calculated $\sigma (n)$  for different values of $T$ with a fixed $s=0.05$ meV for 2D holes in GaAs/AlGaAs quantum wells of width $a=200$ \AA. \ The 2D interface impurity density is $1.5 \times 10^8$ cm$^{-2}$. The 3D remote impurities are distributed randomly from $z_1 = 1600$ \AA \ to $z_2 = 1650$ \AA \  with 3D average density $7 \times 10^{14}$ cm$^{-3}$.  }
\label{fig19}
\end{center}
\end{figure}

(i) Is there a true phase transition in the theory? The answer is ``no" -- our theory is by construction a pure crossover theory where in the ``higher-density" metallic phase diffusive conduction by the mobile carriers dominates whereas in the ``low-density" insulating phase activated conduction by the trapped carriers dominates. But the crossover can be very sharp, and depending on the situation, could easily mimic a sharp metal-to-insulator phase transition.  In Fig.~\ref{fig19} we show our numerical results for $\sigma(n)$ for different values of $T$ for a fixed $s$, clearly showing that the $\sigma(n)$ curves for different temperatures do not go through a single value of critical transition density, but are bunched in a small crossover density regime.

(ii) How does the crossover (or transition) density depend on the disorder strength? The transition density always increases with increasing the potential fluctuation strength ``$s$". Thus, more disordered systems will behave insulating at higher carrier density. This is obvious from the results shown in Sec.~\ref{sec:3}. A corollary of this finding, which is also apparent in the results of Sec.~\ref{sec:3}, is that for larger disorder with higher $s$ values, the temperature dependence of the resistivity in the metallic phase will be strongly suppressed since the temperature dependence of the diffusive $\rho(T)$ for the mobile carriers is determined by the parameters $T/T_F$ and $q_{TF}/k_F$, both of which decrease with increasing density. Thus, very low disorder 2D systems with small values of $s$ are necessary for the experimental observation of the 2D MIT phenomena.

(iii) Is there a temperature dependence of the crossover density? The answers to questions (i) and (ii) and a careful reading of our presented results in Sec.~\ref{sec:3} show that indeed there is a weak temperature dependence of the crossover density -- thus, the crossover density depends both on the disorder strength and the temperature scale of the experiment. In particular, the transition density increases with increasing temperature (and it also increases with increasing disorder). The reason for this is a bit trivial since it arises from the fact that at higher temperatures (and in the absence of phonon scattering) $\rho(T)$ shows monotonically slowly decreasing behavior with increasing temperature upto rather high density, thus making it appear that the metallic phase is pushed upto higher density as temperature increases since activated conduction becomes increasingly important at higher $T$. This trend of an apparent increase of the crossover density with increasing temperature (or equivalently, an apparent decrease of the ``critical" density with decreasing temperature) has been experimentally observed\cite{Hanein_PRBR98Shahar}. A recent experimental work\cite{qiurichard_PRB11} also shows the dominance of activated conduction at higher temperatures leading to a $\rho(T)$ decreasing at higher temperatures consistent with our 2-phase coexistence model.

(iv) What about the in-plane magnetic field dependence of the 2D resistivity? We now comment on an important class of experiments where an in-plane magnetic field is applied parallel to the 2D system\cite{dashwangprlparallel_PRL00,dashwanglw_PRB05,dashwang_simiPRB05} with the consequent finding that the effective metallic (insulating) phase is strongly suppressed (enhanced), i.e. the crossover density increases with increasing in-plane magnetic field. This phenomenon can be explained in our theory simply as a magnetic field induced enhancement of the trapped immobile carrier density in the system through an increase of the potential fluctuation parameter ``$s$" due to the increasing spin polarization of the system (induced by the applied in-plane field). In particular, for a large enough magnetic field, the 2DEG becomes completely spin-polarized, thus reducing the screening wavevector $q_{TF}$ by a factor of 2, which then enhances ``$s$" by a factor of 2, in turn leading to a factor of 2 increase in the trapped carrier density (see Eq.~\eqref{Eq0:de}) even for exactly the same disorder configuration. This factor of 2 increase in the trapped carrier density then leads to a large increase in the crossover density for the metal to the insulator transition. Such an enhancement (roughly by a factor of 2) in the ``critical" density for 2D MIT in the presence of a strong in-plane applied field has been reported in many experiments\cite{yoon_PRL00,pudalov_conden,pillarisetty_PRl03,Tsui_PRB05,lai_PRB05}.

\begin{figure}[htb]
\begin{center}
\includegraphics[width=0.99\columnwidth]{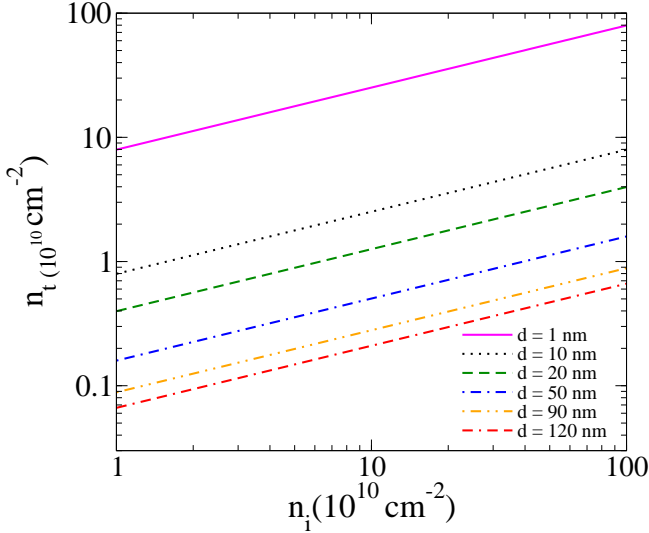}
  \caption{(Color online). Calculated $n_t = \frac{\sqrt{n_i}}{4 \pi z_0}$ as a function of $n_i$.}
\label{fig20}
\end{center}
\end{figure}

(v) What is the relationship between the random charged impurities (defined by $n_i$ and $z_0$) and the potential fluctuations (defined by the variance ``$s$" of the Gaussian probability distribution function $P(V)$ for the disorder-induced potential fluctuations)? In principle, ``$s$" is related to $n_i$, $z_0$ since the random charged impurities give rise to the potential fluctuations. In our theory, we have used a simple minimal model where $n_i$, $z_0$, and $s$ are three independent parameters defining the impurity disorder, but in principle they are related. This relationship is, however, in general unknown since the precise impurity distribution (modeled in our theory by only two parameters $n_i$ and $z_0$) is unknown (and would, in principle, require an infinite number of parameters to define). In addition, even for a given $n_i$ and $z_0$, the real calculation of ``$s$" is numerically intractable and impractical  because one must carry out a nonlinear and self-consistent density functional theory where the carrier density distribution $n(\mathbf{r})$ and the potential fluctuations $V (\mathbf{r})$ determine each other. Within a linear screening theory (which surely does not apply at low carrier density) we find $s = e^2\sqrt{\pi n_i}/(\sqrt{2} \kappa z_0 q_{TF})$ with $q_{TF} = g_s g_v m e^2 /(\kappa \hbar^2)$, and $n_t \equiv \sqrt{n_i}/(4 \pi z_0) = (\frac{g_s g_v m}{2 \pi \hbar^2})(\frac{s}{\sqrt{2 \pi}})$. In Fig.~\ref{fig20} we show our calculated $n_t$ as a function of $n_i$ for 2D GaAs holes for different values of $z_0$. These relationships also tell us how the 2D material (i.e. Si or GaAs, electrons or holes, etc.) enters the theory through the system parameters $m$, $\kappa$, $g_s$, $g_v$. For example, 2D electrons in GaAs quantum wells have $m=0.07 m_e$ in contrast to 2D holes with $m = 0.3 m_e$ with all the other parameters being the same. This implies that for the same disorder parameters the effective ``$s$" in 2D holes is four times smaller than in 2D electrons, leading to much more prominent 2D MIT behavior in 2D GaAs holes than in 2D GaAs electrons in the same carrier density and temperature regime as observed experimentally\cite{manfra2007,lilly2003,dassarma2005b}. By contrast 2D Si electrons have $m=0.19 m_e$ but $g_v = 2$ with a ``$\kappa (\approx 8)$" which is somewhat smaller than in GaAs ($\kappa \approx 12$). Thus, for the same disorder parameters, 2D GaAs holes will manifest somewhat stronger 2D MIT phenomena than 2D Si electrons, as observed experimentally\cite{tracy2009}.

\section{Conclusion}
\label{sec:5}

Our physically motivated 2-component transport theory based on the coexistence of two phases of carriers in the 2D disordered system, trapped/activated carriers and mobile/diffusive carriers, provides a qualitative description of the experimentally observed features of the 2D MIT phenomenology. The basic model of the 2-phase coexistence (``trapped" and ``mobile") is classical, somewhat akin to coexisting solid (``trapped") and liquid (``mobile") phases in the lakes-and-hills fluctuating potential landscape, although we calculate the 2D conductivity using a semiclassical Boltzmann transport theory where carrier scattering is treated quantum-mechanically. In our theory, the basic metal-to-insulator transition is a crossover phenomenon where with lowering carrier density, the relative activated contribution of the trapped carriers to the net conductivity in the 2-component model increases with respect to the diffusive contribution of the mobile carriers, reflecting eventually in an apparent change in sign of $d\rho/dT$ at low temperatures and densities. The overall temperature and density (as well as in-plane magnetic field) dependence of the transport properties arises from a complex interplay among potential fluctuations, screening, and activated transport.

The main effects left out of the theory are quantum tunneling, quantum interference, and electron-electron interaction (beyond RPA screening). We do not see an easy way of including these effects in our  model. Tunneling can be included\cite{meir1999,meir_PRB00} within a Landauer-Buttiker type approach, but then many realistic effects such as screening and diffusive impurity scattering must be left out.  In principle, at $T=0$ all states in 2D systems become localized, but experimentally this limit seems difficult to reach. The advantage of our theory is that one can do concrete microscopic calculations for the density, temperature, and disorder dependent 2D conductivity for different 2D materials. The disadvantage is that it is not easy to see how to include quantum effects such as localization/tunneling and interaction, which must eventually play a role at low enough temperatures.

It may be useful to emphasize two aspects of our theory.  First, the results presented in this work are not only in excellent qualitative agreement of the existing experimental results, they are, in fact, in good quantitative agreement with the experimental 2D MIT data in the literature.  In particular, the theoretically calculated temperature dependence of the 2D metallic resistivity at higher carrier density agrees well with the existing experimental data by virtue of our inclusion of the full temperature- and density-dependent screening of the underlying Coulomb disorder as already discussed extensively in the earlier works \cite{dassarma1999,dassarma2000,dassarma2004,dassarma2003,dashwangprlparallel_PRL00,
dashwanglw_PRB05,dashwang_simiPRB05} where it was established that the strong linear metallic temperature dependence of the 2D metallic resistivity arises from temperature dependence of the 2D screening function around $2 k_F$.  Similarly, the strong exponential insulating temperature-dependence for lower carrier density is guaranteed in our effective medium approximation by the activated conduction of the trapped carriers.  Experimentally, of course, the quantitative details vary enormously from system to system since the relevant disorder parameters (e.g. $n_i$, $z_0$, $s$, etc.) vary strongly among different 2D systems. For example, in Si MOSFETs, $z_0$ is typically around 1 nm whereas $n_i$ is typically $10^{10}$ cm$^{-2}$, which leads to $n_t  \sim 10^{11}$ cm$^{-2}$, which is in good quantitative agreement\cite{abrahams2001,kravchenko1994} with the so-called 2D MIT critical density in Si-based 2D systems.  In modulation-doped 2D GaAs systems, where the background charged impurity density is negligibly small, the main scattering mechanism is provided by the remote dopants, leading to $z_0 \sim  100$ nm or larger with $ n_i \sim 10^{10}$ cm$^{-2}$.  This leads to $n_t  \sim 10^9$ cm$^{-2}$, which is in agreement with the observed critical density\cite{lilly2003,manfra2007} in high-mobility 2D GaAs systems. In Fig.~\ref{fig20}, we explicitly show the dependence of $n_t$ on $n_i$ and $z_0$, which can be directly compared with experiments.  One serious problem in this context is, of course, that the precise values of $n_i$ and $z_0$ are unknown in experimental samples, and therefore, any direct quantitative comparison is not very meaningful (which is the reason for emphasizing qualitative rather than quantitative agreement with the data throughout this paper) since the theory depends sensitively on the precise values of these disorder parameters.  We emphasize, however, that to the extent the disorder parameters in the experimental samples are known, our results provide a reasonable semi-quantitative agreement with the experimental data.  Second, our theory should at best be thought of as a zeroth order semi-classical 2-component description of the reality because of our neglect of many important physical effects invariably present in nature (e.g. Anderson localization, electron-electron interaction, quantum tunneling).  A complete theory including both the effects of long-range Coulomb disorder (the main ingredient included in our theory) and localization/interaction/tunneling is at this stage not only beyond the scope of our work, it is in fact essentially impossible.

Our work is motivated entirely by the belief that the zeroth order physics for 2D MIT is captured well by considering the long-range Coulomb disorder (leading to the model of trapped and mobile carriers) effects and leaving out other effects. All we can say at this stage, other than emphasizing the qualitative agreement between our results and experiments, is that earlier works in many different 2D systems\cite{manfra2007,tracy2009,jiang_apl1533,ilani2000,ilani2001,
leturcq_PRL03,dassarma2005b,allison2006,adam2008,tracy2006,efros1989,
pikus_spj89,nixon1990,efros1993,shi2002,meir1999,meir_PRB00} have already claimed the applicability of a 2-component classical percolation model to the 2D MIT phenomenon, and our work simply carries out a quantitatively complete transport calculation using the 2-component model so that the actual theoretical transport results can be explicitly obtained using the effective medium theory.  More work would obviously be needed to complete the story of 2D MIT phenomenon in figuring out the precise roles played by quantum tunneling Anderson localization, and electron-electron interaction in the transport physics of 2D disordered systems.

A relevant question in this context is what one can learn about the experimental 2D MIT phenomena from our 2-component theoretical analysis and our presented transport results in this work. After all, our theory is explicitly constructed as a crossover theory with both mobile and trapped carrier contributing to the conduction (with their relative contributions changing in a complicated manner depending on disorder parameters, carrier density, and temperature), and as such, our work obviously cannot conclusively theoretically establish 2D MIT to be a crossover phenomenon since we do not compare and contrast our crossover theory to any quantum critical theory treating 2D MIT to be a true quantum localization transition. (We comment as an aside that no detailed theoretical transport results are available in the literature based on a quantum critical model of 2D MIT.) The key message of our analyses is that the 2-component semiclassical percolation theory requires very reasonable system parameters to theoretically reproduce the experimentally observed 2D resistivity $\rho (n, T)$ over a very wide range of carrier density and temperature without any fine-tuning of the generally unknown disorder parameters ($n_i$, $z_0$, and $s$ in our theory for the specific impurity disorder mechanism). In particular, the same set of unique values for ($n_i$, $z_0$, $s$) reproduces the approximate ``critical" carrier density regime for the metal-to-insulator crossover transition in $\rho(T)$ as well as the approximate (and often strong) temperature dependence for $\rho(T)$ in the nominally metallic phase. In addition, the same set of $n_i$, $z_0$, and $s$ (which reproduces the correct crossover density scale) also reproduces the absolute magnitude of $\rho (T,n)$ both in the metallic and in the insulating regime. This is obviously a rather nontrivial theoretical accomplishment indicating that the temperature dependence of $\rho (T, n)$ on the metallic side is closely connected with both the actual density for the metal-to-insulator transition as well as the behavior of $\rho(T)$ on the insulating side. This can be easily inferred by comparing our Fig.~\ref{fig4} with Fig.~\ref{fig1} in the corresponding experimental work\cite{manfra2007}. Both our Fig.~\ref{fig4} and Fig.~\ref{fig1} in Ref.~[\onlinecite{manfra2007}] indicate a crossover carrier density around $3 - 5 \times 10^9$ cm$^{-2}$, and the measured $\rho(n, T)$ in the metallic regime in Ref.~[\onlinecite{manfra2007}] is precisely in the same range as our calculated values in Fig.~\ref{fig4}. The change in $\rho(T)$ on the metallic side agrees well between our theory and the data in Ref.~[\onlinecite{manfra2007}]. In addition, our calculated effective conductivity exponent, $\delta' \approx 1.6$ in Fig.~\ref{fig6} agrees well with the experimentally measured exponent given in Fig.~\ref{fig5} of Ref.~[\onlinecite{manfra2007}]. What is important here is not the precise quantitative agreement, but the fact that very reasonable agreement between several different independent experimental quantities and our theoretical calculations are achieved using a single set of reasonable parameter values for $n_i$, $s$, and $z_0$. This seems to indicate that the qualitative agreement between the 2-component crossover theory and experimental data is unlikely to be a coincidence (since we are not fine tuning unknown parameters), and the experimental 2D MIT phenomenon is likely to be strongly affected by the percolation physics presented in this work. This is the key message of our work.

A direct corollary of the above discussion about the agreement between out percolation theory and 2D MIT experiments (without the fine-tuning of disorder parameters) is that any experiment finding a sharp separatrix (i.e. a sharp value of a temperature-independent critical density $n_c$ corresponding to a sharp temperature-independent critical resistance $\rho_c$) distinguishing the low-density ($n< n_c$) insulating phase from the high-density ($n >n_c$) metallic phase with a characteristic critical resistance $\rho_c \equiv \rho (n=n_c)$ is in fundamental qualitative disagreement with our theoretical approach. If a sharp temperature independent separatrix exists distinguishing the metallic phase from the insulating phase, then our theory is inapplicable and the physics is likely to be a quantum critical phenomenon (and not a crossover as we have assumed in our model). In the early days of 2D MIT physics, such a sharp separatrix and an associated critical density/resistance for 2D MIT phenomenon were often claimed to be true experimentally\cite{kravchenko1994}. But it is now well-established that most 2D MIT experiments do not have any sharp separatrix (or an associated critical resistance), and this was already emphasized by Hanein {\it et al}. as far back as 1998\cite{Henein_PRL98}. In fact, the experimental results presented in Fig.~\ref{fig1} of this paper clearly show that the typical characteristic resistance associated with 2D MIT crossover varies from $\sim 8$ k$\Omega$ in Fig.~\ref{fig1}(a) through $\sim 40$ k$\Omega$ in Fig.~\ref{fig1}(c) to $\sim 50$  k$\Omega$ in Fig.~\ref{fig1}(b), and this ``critical resistance" is strongly temperature dependent in each sample. In the literature, the experimentally observed critical resistance varies by more than an order of magnitude\cite{abrahams2001,kravchenko2004,dassarma2005,Spivak_RMP10,dassarma2010} in different systems and samples, casting doubts on the early claim of a universal $\rho_c$ characterizing 2D MIT. We must emphasize, however, that our theory is simply in applicable to any experiment, such as those reported in Ref.~[\onlinecite{kravchenko1994}], where the 2D MIT is characterized by a sharp separatrix and an associated precise critical resistance. Fortunately (for our theory), many experimental observations are consistent with 2D MIT being a crossover phenomenon with no precise critical density (and critical resistance) including the data shown in Fig.~\ref{fig1} of the current paper. It is of course, possible (although unlikely in our opinion) that some specific subsets of 2D MIT experiments are indeed observing a quantum critical transition and others (e.g. the three distinct experimental results depicted in Fig.~\ref{fig1} of our paper) are observing crossover behaviors. This could, in principle, arise from the underlying effective disorder mechanism controlling 2D MIT transport behavior being different in different systems and samples studied in different experiments. For example, GaAs- based 2D systems could be dominated by long-range Coulomb disorder whereas Si MOSFETs (studied in Ref.~[\onlinecite{kravchenko1994}]) could be dominated by short-range disorder due to the fact that the GaAs 2D samples are modulation doped with the ionized dopants being far from the 2D carriers whereas in Si MOSFETs the main scattering is by nearby impurities in the oxide layer and by interface roughness.

In the unlikely scenario that such a dichotomy exists with some 2D MIT phenomena being quantum critical and others being semiclassical crossover, our theory would apply only to the situation where the transport is dominated by long-range Coulomb disorder (leading to the inhomogeneous potential landscape) and not to the situation dominated by short-range white noise disorder.

Finally, we conclude with some brief remarks on the precise quantitative comparison between our theoretical numerical results and the experimental data of Ref.~[\onlinecite{manfra2007}]: (i) The most important discrepancy is that we overestimate the resistivity deep in the insulating phase where our theoretical resistivity is $\sim 10^7$ ohms whereas the corresponding experimental resistivity in Ref.~[\onlinecite{manfra2007}] is around $\sim 10^6$ ohms. This order of magnitude discrepancy most likely arises from our neglecting the variable range hopping conduction in the insulating phase which would strongly enhance the insulating conductivity. (ii) Our transition density tends to be somewhat smaller ($\sim 10 \%$) than the experimental transition, which again could be due to our neglect of the variable range hopping conduction in the insulating phase. (iii) Typically, the calculated resistivity in our Figs.~\ref{fig7} and \ref{fig13} is somewhat (by about $30 \%$ or so) smaller than the experimental data at higher carrier density, which may be due to our neglect of weak localization corrections. (iv) On the other hand, our calculated resistivity in Figs.~\ref{fig4}, \ref{fig10} and \ref{fig16} is somewhat larger (by about $50 \%$) than the experimental data. (v) The last two points together indicate that a complete model of disorder, which includes all different disorder mechanisms together (i.e. interface disorder, remote disorder, bulk disorder) that we have treated separately in our numerical results, may very well be able to reproduce the experimental results precisely quantitatively (except for the strongly insulating resistivity discussed in item (i) above). But such a theory would have at least ten independent disorder parameters, and getting quantitative agreement by adjusting ten independent disorder parameters would essentially be a meaningless data-fitting exercise. Without more quantitative information about the underlying disorder, our current theory and results are probably the best one can do.

\begin{acknowledgments}

This work is supported by LPS-CMTC and Microsoft Q.

\end{acknowledgments}

\end{document}